\magnification1095
\input amstex
\hsize32truecc
\vsize44truecc
\input amssym.def
\font\ninerm=plr9 at9truept
\font\twbf=cmbx12
\font\twrm=cmr12

\footline={\hss\ninerm\folio\hss}

\def\section#1{\goodbreak \vskip20pt plus5pt \noindent {\bf #1}\vglue4pt}
\def\eq#1 {\eqno(\text{\rm#1})}
\let\al\aligned
\let\eal\endaligned
\let\ealn\eqalignno
\let\dsl\displaylines

\let\o\overline
\let\ul\underline
\let\a\alpha
\let\b\beta
\let\d\delta
\let\e\varepsilon
\let\f\varphi
\let\D\Delta
\let\g\gamma
\let\G\varGamma
\let\P\varPsi
\let\F\varPhi
\let\la\lambda

\let\z\zeta
\def\gh{{\frak h}}
\def\gg{{\frak g}}
\def\gm{{\frak m}}

\let\t\widetilde
\let\u\tilde

\def\2{^{\text{\rm II}}}

\def\uP{\ul{\t P}{}}
\def\hG{\widehat{\overline\G}}
\def\RG{\t R(\t\G)}
\def\({\left(}\def\){\right)}
\def\[{\left[}\def\]{\right]}
\def\dr#1{_{\text{\rm#1}}}

\def\ct{constant}
\def\co{cosmological}

\def\il{inflation}

\def\q{quintessence}

\def\KK{Kaluza--Klein}
\def\JT{Jordan--Thiry}
\def\nos{nonsymmetric}
\def\up#1{\uppercase{#1}}
\def\eu{\expandafter\up}
\def\dint{-\kern-11pt\intop}
\def\wiel{2\mu^3+7\mu^2+5\mu+20}
\def\lon{\ln\(|\z|+\sqrt{\z^2+1}\)+2\z^2+1}

{\twbf
\advance\baselineskip4pt
\centerline{M. W. Kalinowski}
\centerline{\twrm (Higher Vocational State School in Che\l m, Poland)}
\centerline{A short note}
\centerline{on a total amount of \il}
}

\vskip20pt plus5pt

{\bf Abstract.} 
We calculate a total amount of an \il\ during two de Sitter phases in our
\co\ models and corresponding masses of \q\ particles.

\vskip20pt plus5pt

In the paper we consider a total amount of an \il\ in \co\ models developed
in Ref.~[1]. The full treatment of the \eu\nos\ \KK\ (\JT) Theory is given in
Ref.~[2]. In our approach we get two de Sitter phases governed by two 
different Hubble \ct s $H_0$ and~$H_1$.
We underline in~[1] that we have to do with phase transitions: of the second
order in the configuration of Higgs' field and of the first order in the
evolution of the Universe.

The Hubble \ct\ in the first phase is given by the formula ($\hbar=c=1$)
$$
\al
H_0&=\frac
{\(-n\g+\sqrt{n^2\g^2+4(n^2-4)\a_1\b}\)^{(n-2)/4}}
{2^{(n+2)/4}(n+2)^{(n+2)/4}\sqrt3\b^{n/4}}\cr
&\qquad{}\times\(n^2\g^2-\g\sqrt{n^2\g^2+4(n^2-4)\a_1\b}+4\a_1\b(n+2)\)^{1/2},
\eal
\eq1
$$
where
$$
\ealn{
\g&=\frac{m^2_{\u A}}{\a_s^2}\uP=\frac1{r^2}\uP &(2)\cr
\b&=\frac{\a_s^2}{l\dr{pl}^2}\t R(\t\G)={\a_s^2}m\dr{pl}^2\t R(\t\G) &(3)\cr
\a_1&=\frac{\a_s^2}{r^2}\(\frac{\a_s l\dr{pl}}r\)^2 \(\frac A{\a_s^6}\)=
m^2_{\u A}\(\frac{m_{\u A}}{m\dr{pl}}\)^2 \(\frac A{\a_s^6}\). &(4)
}
$$
$\a_s$ is a coupling \ct, $m\dr{pl}$ and $l\dr{pl}$ are Planck's mass and
Planck's length, respectively, $m_{\u A}$ and~$r$ are a scale of an energy (a
scale of a mass of broken gauge bosons). 

$\t R(\t\G)$ is a scalar of a curvature of a connection on a group
manifold~$H$. 
$$
\uP=\frac1{V_2}\intop_M R(\hG)\sqrt{|\t g|}\,d^{n_1}x, \eq5
$$
where $V_2$ is a measure of a manifold $M$ (a vacuum states manifold),
$R(\hG)$ is a scalar of a curvature for a connection defined on~$M$, $\t
g=\det(g_{\u a\u b})$ (see~[2]).
$$
A=\frac1{V_2}\intop_M
\(l_{ab}\(g^{[\u m\u n]} g^{[\u a\u b]} C^b_{\u a\u b} C^a_{\u m\u n}
-C^b_{\u m\u n}g^{\u a\u n}g^{\u b\u n}\u L^a_{\u a\u b}\)\)
\sqrt{\u g}\,d^{n_1}x,  \eq6
$$
$$
l_{dc}g_{\u m\u b}g^{\u c\u b}\t L^d_{\u c\u a}+l_{cd}g_{\u a\u m}g^{\u m\u c}
\t L^d_{\u b\u c}=2l_{cd}g^{\u a\u m}g^{\u m\u c}C^d_{\u b\u c}. \eq7
$$
$C^d_{\u b\u c}$ are structure \ct s for a Lie algebra $\gh$ (a Lie algebra
of a group~$H$).

Using Eqs (2), (3), (4) one gets
$$
\al
H_0&=\(\frac{m_{\u A}}{\a_s}\)\(\frac{m_{\u A}}{\a_s^2m\dr{pl}}\)^{n/2} \frac
{\(-n\uP+\sqrt{n^2{\uP}^2+4(n^2-4)\t R(\t\G)A}\)^{(n-2)/4}}
{2^{(n+2)/4} (n+2)^{(n+2)/4}\sqrt3({\t R}(\t\G))^{n/4}}\cr
&\qquad{}\times\(n^2{\uP}^2-\uP\sqrt{n^2{\uP^2}+4(n^2-4)A\t R(\t\G)}
+4(n+2)A\t R(\t\G)\)^{1/2},
\eal
\eq1a
$$
In the case of more general models (two different types of critical points
for Higgs' field configuration) we take for $A$
$$
A=4\a_s^2V(\F\dr{crt}^1) \eqno(*)
$$
where $V(\F\dr{crt}^1)$ is a value of Higgs' potential for $\F\dr{crt}^1$
(a~metastable state Higgs' field configuration).

In the case of a Hubble \ct\ $H_1$ we get
$$
H_1=\frac{|\g|^{(n+2)/4}n^{n/4}}{\sqrt3 \b^{n/4} (n+2)^{(n+2)/4}} \eq8
$$
or
$$
H_1=\(\frac{m_{\u A}}{m\dr{pl}}\)^{n/2}\frac{m_{\u A}}{\a_s^{n+1}}\,
\frac{|\uP|^{(n+2)/4}n^{n/4}}{\sqrt3  (n+2)^{(n+2)/4}(\t R(\t \G))^{n/4}} \eq8a
$$
In this case we suppose that $\g<0$ and $\a_1, \b>0$. This is possible because
$\g$ is a function of a \ct~$\zeta$ and can have any sign (see Ref.~[2]). In
the case of~$\b$ it is the same. $\b$~is a function of a \ct~$\xi$ ($\mu$~in
Ref.~[2]). This is because of properties of $\t R(\t\G)$ and $\uP$.

In Ref.\ [3] we consider several consequences of Higgs' field dynamics on
primordial fluctuations (perturbations) spectrum. We find three different
functions and examine their properties. We write down also an equation for an
amount of an \il\ in the first de Sitter phase for these functions (see Refs
[1],~[3]). They are calculable for functions (9) and~(82) (see Ref.~[3]). In
the case of a function~(64) (see Ref.~[3]) we get under some practical
assumption 
$$
N_0=1.96\,\frac{H_0}b=1.96\,\frac{6r^2H_0^2}{e^{n\P_1}A}=
\frac{1.96}{\a_s^4}\(\frac{m_{\u A}}{m\dr{pl}}\)^2 F(\b,\g,\a_1)
\eq9
$$
where $F(\b,\g,\a_1)$ is given by the formula
$$
F(\b,\g,\a_1)=
\frac{\(n^2\g^2-\g\sqrt{n^2\g^2+4(n^2-4)\a_1\b}+4\a_1\b(n+2)\)}
{(n+2)\a_1\(-n\g+\sqrt{n^2\g^2+4(n^2-4)\a_1\b}\)}
\eq10
$$
or
$$
N_0=1.96\,\(\frac{m_{\u A}}{m\dr{pl}}\)^4
F(\t R(\t\G),\uP,A). \eq10a
$$

Now let us come to the second de Sitter phase. During this phase the \il\ is
driven by the field~$\P$ which changes (slowly $\dot\P\approx0$) from $\P_0$
to $\frac12\ln\(\frac\b{|\g|}\)$. In Ref.~[3] we develop two approximation
schemes for an evolution of~$\P$. In the case of a harmonic oscillation we
calculate an amount of an \il. We find also a different scheme---the
so called slow-roll approximation which offers an infinite in time evolution
of~$\P$. 

Moreover from practical point of view we can suppose that an evolution
starts if $\P$ is very closed to $\frac12\ln\(\frac{n|\g|}{(n+2)\b}\)$. In
term of a variable 
$$
y=\sqrt{\frac\b{|\g|}}\,e^\P \eq11
$$
$y$ is closed to $\sqrt{\frac n{n+2}}$. Let $y$ be
$$
y=\sqrt{\frac n{n+2}}+\e \eq12
$$
where $\e>0$ is very small. Let us take 
$$
\e=\frac1{n(n+1)(n+2)}\,. \eq13
$$
We remind that $n\ge14$. Thus 
$$
\e<3\cdot10^{-4}. \eq14
$$

Let us consider the formula (3.259) from Ref.~[1] and let us calculate the
limit 
$$
\lim_{y\to \sqrt{\frac n{n+2}}+\e}  I. \eq15
$$
One gets after some simplifications (taking under consideration the fact that
$n>14$)
$$
\lim_{y\to \sqrt{\frac n{n+2}}+\e}I=
-\frac\pi2-\frac1{2\sqrt{2n}}\ln2. \eq16
$$
From the other side we have
$$
\lim_{y\to1}I=-\frac\pi2+\frac1{\sqrt{2n}}\ln\(n+1+\sqrt{n(n+2)}\). \eq17
$$
From Ref.~[1] we have
$$
I=\o B(t-t_0) \eq18
$$
where 
$$
\o B=4\sqrt{2\pi}(n+2)|\g|\(\frac n{n+2}\)^{n/2}\(\frac{|\g|}\b\)^{n/2}
\frac1{m\dr{pl}}\sqrt{\o M} \eq19
$$
or
$$
\o B=4\sqrt{2\pi}\(\frac{m_{\u A}}{m\dr{pl}}\)^{n+1} (n+2) 
\(\frac n{n+2}\)^{n/2}
\frac{m_{\u A}}{\a_s^{2(n+1)}} |\uP|
\(\frac{|\uP|}{\RG}\)^{n/2}\frac1{\sqrt{\o M}}\,.
\eq19a
$$
Writing
$$
\ealn{
I(1)&=\o B(t\2\dr{end}-t_0) &(20)\cr
I\(\sqrt{\frac n{n+2}}+\e\)&=\o B(t\2\dr{initial}-t_0) &(21)
}
$$
where $t\2\dr{initial}$ and $t\2\dr{end}$ mean an initial and end time of the
second de Sitter phase, one gets
$$
\D t\2=t\2\dr{end}-t\2\dr{initial}=
\frac1{\,\o B\,}\,\frac1{\sqrt{2n}}\ln\(\sqrt2\(n+1+\sqrt{n(n+2)}\)\) \eq22
$$
where $\D t\2$ means a period of time of the second de Sitter phase. Thus the
amount of \il\ for this phase is
$$
N_1=H_1\D
t\2=\frac{H_1}{\,\o B\,}\,\frac{\ln\(\sqrt2\(n+1+\sqrt{n(n+2)}\)\)}{\sqrt{2n}}\,. \eq23
$$
 
Using formulae (19) and (8) one gets
$$
N_1=\sqrt{2\pi}\(\frac\b{|\g|}\)^{n/4}\(\frac{n+2}n\)^{n/4}
\frac{\ln\(\sqrt2\(n+1+\sqrt{n(n+2)}\)\)}{\sqrt{6n|\g|}(n+2)^3}
\cdot \frac{m\dr{pl}}{\sqrt{\o M}} \eq24
$$
or
$$
\al
N_1&=\(\frac{2\sqrt{2\pi}}{\sqrt{\o M}}\)\(\frac{m\dr{pl}}{m_{\u A}}\)^{(n+2)/2}
\(\frac{\RG}{|\uP|}\)^{n/4}\(\frac{n+2}n\)^{n/4}\cr
&\times\frac{\ln\(\sqrt2\(n+1+\sqrt{n(n+2)}\)\)}{2\sqrt6 \a_s^{n-1} |\uP|^{1/2}(n+2)^3}
\,.
\eal \eq24a
$$

For large $n$ one finds
$$
N_1\simeq \sqrt{2\pi} \sqrt{\frac e6} \(\frac\b{|\g|}\)^{n/4}
\frac{\ln n}{|\g|^{1/2}n^{7/2}} \cdot \frac{m\dr{pl}}{\sqrt{\o M}} \eq25
$$
or 
$$
N_1\simeq \sqrt{\frac e6} \(\frac{\RG}{\uP}\)^{n/4}
\(\frac{m\dr{pl}}{m_{\u A}}\)^{(n+2)/2}\(\frac{\sqrt{2\pi}}{\sqrt{\o M}}\)
\frac{\ln n}{\a_s^{n-1}|\uP|^{1/2}n^{7/2}}\,. \eq25a
$$
The total amount of an \il\ considered in the paper is
$$
N\dr{tot}=N_0+N_1 \eq26
$$
and should be fixed to ${}\sim60$.

In order to give an example of these calculations let us consider a
six-dimensional Weinberg-Salam model (see Ref.~[4]). It is of course a
bosonic part of this model. In this case $H=G2$, $\dim H=14$, $M=S^2$
(two-dimensional sphere). This will be of course the model in \eu\nos\ \KK\
(\JT) Theory. In order to simplify the calculations we take
$$
\t R(\t \G)=\frac{2(2\mu^3+7\mu^2+5\mu+20)}{(\mu^2+4)^2} \eq27
$$
(see formula (7.21) from the second point of Ref.~[2]).

The scalar curvature has been calculated here for $H=SO(3)\simeq SU(2)$.
Moreover it gives a taste of the full theory
$$
\al
&\uP=\uP(\z)\cr
&=\biggl\{\frac{16|\z|^3(\z^2+1)}{3(2\z^2+1)(1+\z^2)^{5/2}}
\(\z^2E\(\frac{|\z|}{\sqrt{\z^2+1}}\)-(2\z^2+1)K\(\frac{|\z|}{\sqrt{\z^2+1}}\)\)\cr
&+8\ln\(|\z|\sqrt{\z^2+1}\)
+\frac{4(1+9\z^2-8\z^4)|\z|^3}{3(1+\z^2)^{3/2}}\biggr\}
\biggm/ \(\ln\(|\z|+\sqrt{\z^2+1}\)+2\z^2+1\) 
\eal
\eq28
$$
(see formula (5.6.64) from the first point of Ref.~[2]).

For $V_2$ one gets
$$
V_{2} = \frac{2\pi}{|\z|}\(\ln\(|\z|+\sqrt{\zeta ^2+1}\)
+ 2\zeta ^{2} + 1\)\eq29
$$
(see formula (5.6.65) from the first point of Ref.~[2]), where
$$\eqalignno{
K(k) &= \intop_{0}^{\pi/2}\frac{d\theta}{\sqrt{1-k^2\sin^2 \theta}}\,,&(30)\cr
E(k) &= \intop_0^{\pi/2}\sqrt{1-k^2\sin^2 \theta }\,d\theta.&(31)
}
$$

$\uP$ has been calculated for an Einstein-Kaufmann connection defined
on~$S^2$. We need $\uP<0$, thus in our case
$$
|\z|>\z_0=1.36\dots \eq32
$$
(see p.~383 and Fig.~15 from the first point of Ref.~[2]).

In order to write the formula in this case we need to calculate $A$ from
Eq.~(6). We should calculate $\t L^d_{\u b\u c}$ from Eq.~(7). In our case
this is quite easy for $S^2$ is 2-dimensional and $\t a,\t b=1,2$ ($\t a,
\t b=5,6$ if we embed $S^2$ as a vacuum state manifold in the full theory). 

In this
case one easily finds
$$
\t L^c_{\u a\u b}=h^{ce}l_{ed}C^d_{\u a\u b} \eq33
$$
where $C^d_{\u a\u b}$ are structure \ct s of the group $H$ in such a way
that $\t a,\t b$ correspond to the complement $\gm$, $\gg=\gg_0 \mathrel{\dot+} \gm$,
$$
\gg \subset \gh \eq34
$$
where $\gg=A_1$, $l_{ab}=h_{ab}+\mu k_{ab}$, $h^{ef}h_{fd}=\d^e_d$.
Using the exact form of the \nos\ tensor on $SO(3)$ and the \nos\ tensor
on~$S^2$ (see formulae (2.2.24a), (5.4.31) from the first point of Ref.~[2])
one gets
$$
\al
\o A&=l_{ab}\(g^{[\u m\u n]}g^{[\u a\u b]}C^a_{\u m\u n}C^b_{\u a\u b}
-C^b_{\u m\u n}g^{\u a\u n}g^{\u b\u m}\t L^a_{\u a\u b}\)\cr
&=-\(\frac{\z^2}{(1+\z^2)^2\sin^2\theta} 
+ \frac{4+\mu^2\sin^2\theta}{(1+\z^2)\sin^2\theta}\) 
\eal \eq35
$$
and finally
$$
\ealn{
A&=\frac1{V_2}\intop_{S^2} \o A\sqrt{\t g}\,d\theta\,d\f=
\frac1{V_2}\intop_0^{2\pi} \intop_0^\pi \,d\f\,d\theta\,\o A\sqrt{\t g}, &(36)\cr
A&=\frac{|\z|\[(5\z^2+4)\ln 2 - 2\mu^2(1+\z^2)\]}
{(1+\z^2)\(\ln\(|\z|+\sqrt{\z^2+1}\)+2\z^2+1\)}&(37)
}
$$
where the integral $\int_0^\pi \frac{d\theta}{\sin\theta}$ has been
calculated in the sense of a principal value:
$$
\intop_0^\pi \frac{d\theta}{\sin\theta}=\lim_{\e\to0^+}
\intop_\e^{\pi-\e}\frac{d\theta}{\sin\theta}=-\ln2. \eq38
$$
Using Eqs (27), (28), (37) one writes the formula (1a) in the form ($n=14$)
$$
\al
H_0&=\frac1{2^{19}\sqrt3}\(\frac{m_{\u A}}{\a_s^2m\dr{pl}}\)^7
\cdot \frac{m_{\u A}}{\a_s}\cr
&\times\frac{\(7g(\z,\mu)+\sqrt{49g^2(\z,\mu)+384h(\z,\mu)}\)^3(\mu^2+4)^3}
{\(\ln\(|\z|+\sqrt{\z^2+1}\)+2\z^2+1\)\(2\mu^3+7\mu^2+5\mu+20\)^{7/2}}\cr
&\times\(98g^2(\z,\mu)+(\mu^2+4)g(\z,\mu)\sqrt{49g^2(\z,\mu)+384h(\z,\mu)}
+32H(\z,\mu)\)^{1/2}
\eal
\eq39
$$
where
$$
\ealn{
g(\z,\mu)&=-f(\z)(\mu^2+4) &(40)\cr
h(\z,\mu)&=|\z|(2\mu^3+7\mu^2+5\mu+20)
\((5\z^2+4)\ln2-2\mu^2(1+\z^2)\)\cr
&\times\frac{\(\ln\(|\z|+\sqrt{\z^2+1}\)+2\z^2+1\)}{(1+\z^2)}&\text{(41a)}\cr
H(\z,\mu)&=h(\z,\mu)\(\ln\(|\z|+\sqrt{\z^2+1}\)+2\z^2+1\)&\text{(41b)}\cr
f(\z)&=\frac{16|\z|^3(\z^2+1)}{3(2\z^2+1)(1+\z^2)^{5/2}}
\(\z^2E\(\frac{|\z|}{\sqrt{\z^2+1}}\)-(2\z^2+1)K\(\frac{|\z|}{\sqrt{\z^2+1}}\)\)\cr
&+8\ln\(|\z|\sqrt{\z^2+1}\)
+\frac{4(1+9\z^2-8\z^4)|\z|^3}{3(1+\z^2)^{3/2}}\,.&(42)
}
$$

From Eq.\ (8a) one gets
$$
\al
H_1&=\sqrt{\frac23}\(\frac{m_{\u A}}{m\dr{pl}}\)^7
\frac{m_{\u A}}{\a_s^{15}}\cdot\frac{7^7}{2^{13}}\cr
&\times\frac{g^4(\z,\mu)(\mu^2+4)^4}{\(2\mu^3+7\mu^2+5\mu+20\)^{7/2}}
\cdot\(\ln\(|\z|+\sqrt{\z^2+1}\)+2\z^2+1\)^{-4}. 
\eal \eq43
$$
From Eq.\ (10a) one finds
$$
\al
N_0&=0.12\(\frac{m_{\u A}}{m\dr{pl}}\)^{4}\cr
&\times\frac{\(7g(\z,\mu)+\sqrt{49g^2(\z,\mu)+384h(\z,\mu)}\)^{-1}(\mu^2+4)^2
(1+\z^2)}
{|\z|\((5\z^2+4)\ln2-2\mu^2(1+\z^2)\)}\cr
&\times\(98g^2(\z,\mu)+(\mu^2+4)g(\z,\mu)\sqrt{49g^2(\z,\mu)+384h(\z,\mu)}
+32H(\z,\mu)\).
\eal
\eq44
$$

Equation (19a) is transformed into
$$
\al
\o B&=\(\frac{m_{\u A}}{m\dr{pl}}\)^{15} \(\frac {7^7
m_{\u A}}{2^{16}\a_s^{30}}\) \(\frac{|f(\z)|}{2\mu^3+7\mu^2+5\mu+20}\)^7
\cdot\frac1{\sqrt{|\o M|}}\cr
&\times \frac{|f(\z)|(\mu^2+4)^6}{\(\ln\(|\z|+\sqrt{\z^2+1}\)+2\z^2+1\)^8}\,.
\eal
\eq45
$$
And finally Eq.\ (24a) gives
$$
\al
N_1&\simeq 5.47\cdot 10^{4}
\(\frac{m\dr{pl}}{m_{\u A}}\)^9
\(\frac{|f(\z)|}{2\mu^3+7\mu^2+5\mu+20}\)^{7/2}\cdot \frac{\a_s}{\sqrt{|\o M|}}\cr
&\times
\frac{\(\ln\(|\z|+\sqrt{\z^2+1}\)+2\z^2+1\)^{3/2}}{|f(\z)|^{1/2}(\mu^2+4)^3}
\,.
\eal
\eq46
$$

Let us notice that according to our assumption $f(\z)<0$ ($\uP<0$), 
$2\mu^3+7\mu^2+5\mu+20>0$ ($\t R(\t\G)>0$), and
$$
\frac{(5\z^2+4)\ln 2}{2(1+\z^2)}>\mu^2 \eq47
$$
(from $A>0$). 

The function on the left hand side of the inequality (47) is rising for
${\z>0}$ and falling for ${\z<0}$, having a minimum at ${\z=0}$ equal to
$2\ln2$. For $\z\to\pm\infty$ it is going to $\frac52 \ln2$. For a specific
value $\z=\z_0=\pm1.36$ (see Eq.~(32)), it is equal to $1.61$. Thus we get
$$
|\mu|\le \sqrt{2\ln2}\approx 1.17741. \eq48
$$
Moreover the polynomial $W(\mu)=2\mu^3+7\mu^2+5\mu+20$ possesses one real root
$$
\mu_0=-\frac{\root3\of{1108+3\sqrt{135645}}}{6}-\frac76-\frac{19}
{6\cdot\root3\of{1108+3\sqrt{135645}}}=-3.581552661\ldots.
$$ 
It is interesting to notice that $W(-3.581552661)=2.5\cdot10^{-9}$ and for
70-digit approximation of~$\mu_0$, $\t \mu$ equal to
$$
-3.581552661076733712599740215045436907383569800816123632201827285932446,
$$
we have
$$
W(\t\mu)=0.1\cdot 10^{-67}.
$$ 

Thus we can
have $\t R(\t\G)>0$ for $\mu>\mu_0$ and in the region given by~(48).
Let us notice that taking sufficiently big $|\z|$ we can make $N_0$ in~(44)
arbiratrily big, i.e.\ about~60. From the other side if we take $|\z|$
sufficiently big, $N_1$ can also be arbitrarily large (i.e.\ $\sim60$). Thus
it seems that in this simple example it is enough to consider arbitrarily
big~$\z$ in order to get large amount of an \il. 

It is interesting to find in this simplified model a \ct~$a$ from Ref.~[1]
(see Eq.~(6)).
One gets after some algebra
$$
\al
a^2&=\frac{\a_s^4B}{Am_{\u A}^2e^{n\P_1}}=
\(\frac{m\dr{pl}}{m_{\u A}}\)^{14}
2^{35}\a_s^{32}\(\frac B{m_{\u A}^2}\)\cr
&\times \(\frac{2\mu^3+7\mu^2+5\mu+20}
{(\mu^2+4)\(7g(\z,\mu)+\sqrt{49g^2(\z,\mu)+384h(\z,\mu)}\)}\)^7\cr
&\times \frac{(1+\z^2)\(\ln\(|\z|+\sqrt{\z^2+1}\)+2\z^2+1\)^8}
{|\z|\((5\z^2+4)\ln2-2\mu^2(1+\z^2)\)}
\eal
\eq49
$$
(we put $n=14$).

The condition 
$$
0<a^2<0.09703 \eq50
$$
gives a constraint on an integration \ct~$B$. In this case $A>0$. However we
consider in Ref.~[1] a special type of a dynamics of Higgs' field with $A<0$.
In this case we have the following constraint imposed on the \ct~$\o a$.
One gets after some algebra
$$
\al
\o a&=\frac{e^{n\P_1}m_{\u A}^2A}{2H_0^2\a_s^2}\cr
&=\frac{192\(7g(\z,\mu)
+\sqrt{49g^2(\z,\mu)+384h(\z,\mu)}\)\(\ln\(|\z|+\sqrt{\z^2+1}\)+2\z^2+1\)}
{98g^2(\z,\mu)+(\mu^2+4)g(\z,\mu)\sqrt{49g^2(\z,\mu)+384h(\z,\mu)}
+32H(\z,\mu)}\cr
&\times\frac{|\z|(\mu^2+4)\((5\z^2+4)\ln2-2\mu^2(1+\z^2)\)}{1+\z^2}\,.
\eal
\eq51
$$
Moreover in this case we should put
$$
\o a = -\frac4{15}, \eq52
$$
i.e.\ $A<0$ and we get
$$
\al
&720 \(7g(\z,\mu)
+\sqrt{49g^2(\z,\mu)+384h(\z,\mu)}\)\(\ln\(|\z|+\sqrt{\z^2+1}\)+2\z^2+1\)\cr
&\quad{}\times |\z|(\mu^2+4)\(2\mu^2(1+\z^2)-(5\z^2+4)\ln2\)\cr
&=\(98g^2(\z,\mu)+(\mu^2+4)g(\z,\mu)\sqrt{49g^2(\z,\mu)+384h(\z,\mu)}
+32H(\z,\mu)\)(1+\z^2).
\eal
\eq53
$$

Eq.\ (53) gives a constraint among parameters of the theory. If Eq.~(53) is
satisfied, we have the dynamics of Higgs' field described by Eq.~(14.381) from the
fifth point of Ref.~[2].

In this way we can consider the spectral function (82) from Ref.~[1] and all
the consequences coming from it. We need of course some supplementary
conditions 
$$
\frac12\cdot\frac{5\z^2+4}{1+\z^2}\cdot \ln2<\mu^2 \eq54
$$
and as usual
$$
\dsl{
\hfill f(\z)<0 \hfill (55)\cr
\hfill 2\mu^3+7\mu^2+5\mu+20>0. \hfill(56)
}
$$
In the case when condition (50) is satisfied we can consider a different
dynamics of Higgs' field described by Eq.~(1) from Ref.~[1] leading to the
spectral function (9) from Ref.~[1] with all the consequences of this
function. In this case we have condition (47) and conditions (55) and~(56).
They are easily satisfied.

The programme of research given in Ref.~[1] will give an additional
constraint and could (in principle) lead to the realistic theory with more
complicated groups and patterns of symmetry breaking. 

Let us notice that in our theory we get \co\ terms. These terms are
described by \ct s $\b,\g$ and~$\a_1$ which are proportional to $\t R(\t\G)$,
$\uP$ and~$A$. The importance of a \co\ \ct\ is now obvious. Thus it is
necessary to control these terms. They depend on \ct s $\z$ and~$\mu$. The
first step in order to control them is to find conditions when they are equal
to zero. In the simplified model we have found these conditions:
$$
\ealn{
\b(\mu_0=-3.581552661\ldots)&=0 &(57)\cr
\g(\z=\pm1.36\ldots)&=0 &(58)
}
$$
and
$$
\a_1=0 \eq59
$$
for
$$
\mu=\pm\sqrt{\frac{\ln2(5\z^2+4)}{2(\z^2+1)}}\,. \eq60
$$

In this way we control the sign of $A$, $\uP$ and $\RG$ playing with \ct s
$\z$ and~$\mu$:
$$
\ealn{
\RG>0 \quad &\hbox{for }\mu>\mu_0=-3.581552661\ldots &(61)\cr
\uP<0 \quad &\hbox{for }|\z|>1.36. &(62)
}
$$
The sign of $A$ is also controllable. All of these results give us
interesting \co\ consequences.

It is interesting to find some conditions on stability of the first de Sitter
evolution of the Universe. In Ref.~[1] we find a criterion
$$
M>M_0 \eq63
$$
where
$$
\al
&M_0=\frac43\cr
&\times\frac{(n+2)\g\sqrt{n^2\g^2+4(n^2-4)\a_1\b}-4(n+2)^2(n-1)\a_1\b
-n^2(n+2)\g^2}
{n\g^2+4(n+2)\a_1\b-\g\sqrt{n^2\g^2+4(n^2-4)\a_1\b}} 
\eal \eq64
$$
or
$$
\al
&M_0=\frac43\cr
&\times\frac{(n+2)\uP\sqrt{n^2\uP^2+4(n^2-4)A\RG}-4(n+2)^2(n-1)A\RG
-n^2(n+2)\uP^2}
{n\uP^2+4(n+2)A\RG-\uP\sqrt{n^2\uP^2+4(n^2-4)A\RG}}\,, 
\eal \eq64a
$$
If
$$
\uP>0,\ \RG>0,\ A>0, \eq65
$$
then
$$
M_0>0. \eq66
$$
Moreover if
$$
\uP<0,\ \RG>0,\ A>0, \eq67
$$
the stability condition is ($M_0<0$)
$$
M<0. \eq68
$$

Moreover for $A<0$ the condition $M_0<0$ is not trivial and we should have
$$
4(n+2)^2(n-1)|A|\RG> 
(n+2)|\uP|\(n^2|\uP|+\sqrt{n^2\uP^2-4(n^2-4)|A|\RG}\). \eq69
$$
In the case of $SO(3)$ group and $S^2$ vacuum states manifold (with $n=14$)
we have
$$
M(SO(3))=-\frac{2(36+7\mu^2)}{(4+\mu^2)M^2\dr{pl}}<0 \eq70
$$
(see the equation on p.~260 of the first point of Ref.~[2]).

Thus if $M_0<0$ the first de Sitter evolution of the Universe is stable. It
means it happens if condition (69) is satisfied. One finds:
$$
\al
&\frac{416|\z|}{1+\z^2}\bigl(2\mu^2(1+\z^2)-(5\z^2+4)\ln2\bigr)\cdot
(2\mu^3+7\mu^2+5\mu+20)\cr
&\qquad{}>\frac{g(\z,\mu)\(49g(\z,\mu)+\sqrt{49g^2(\z,\mu)-384|h(\z,\mu)|}\)}
{\ln\(\(|\z|+\sqrt{\z^2+1}\)+2\z^2+1\)}\,.
\eal \eq71
$$
We have also supplementary conditions
$$
\ealn{
&|\z|>|\z_0|=1.36\ldots&(72)\cr
&\mu>\mu_0=-3.581552661\ldots &(73)\cr
&\mu^2>\frac{\ln2}2\cdot\frac{5\z^2+4}{\z^2+1} &(74)
}
$$

During the second de Sitter phase the evolution of the Universe is unstable
against small perturbation of initial data. Moreover it can be made stable if
$M<0$. However, the second de Sitter phase in our model should be unstable
because it ends with a radiation epoch. The instability of this phase in this
particular case is caused by some physical processes beyond \co\ models.

The interesting point in the theory is a mass of the scalar field (a scalar
particle) during both de Sitter phases. However, in that case we should
consider a \q\ field~$Q$
$$
Q=\sqrt{|\o M|}\frac\P{M\dr{pl}}\,. \eq75
$$
In both de Sitter phases we can consider small oscillations $q_k$ of the \q\
field around an equilibrium
$$
Q=Q_k+q_k, \qquad k=0,1. \eq76
$$
For these small oscillations one finds
$$
{\o m}_k^2=-\frac12 \, \frac{d^2\la_{ck}}{d\P^2}(\P_k) \eq77
$$
or
$$
\al
{\o m}^2_1&=\frac
{-\(-n\g+\sqrt{n^2\g^2+4(n^2-4)\a_1\b}\)^{(n-2)/2}}
{2^{(n+4)/2}(n+2)^{n/2}\b^{n/2}}\cr
&\times \(n\g\sqrt{n^2\g^2+4(n^2-4)\a_1\b}-2\g^2n^2
-4(n+2)(n-3)\a_1\b\)
\eal
$$
$$
\al
&=\(\frac{m_{\u A}}{\a_sm\dr{pl}}\)^n\(\frac{m_{\u A}}{\a_s}\)^2
(n+2)^{-n/2}2^{-(n+4)/2}\cr
&\times\(-\uP+\sqrt{n^2\uP^2+4(n^2-4)A\t R(\t\G)}\)^{(n-2)/2}\cdot
\frac1{\(\t R(\t\G)\)^{n/2}}\cr
&\times\(2n^2\uP^2+4(n+2)(n-3)A\t R(\t\G)
-n\uP\sqrt{n^2\uP^2+4(n^2-4)A\t R(\t\G)}\)
\eal
\eq78
$$
$$
\al
m_0^2&=\frac12 n|\g|\(\frac n{n+2}\)^{n/2}\(\frac{|\g|}{\b}\)^{n/2}\cr
&=\frac n2 \cdot \(\frac{m_{\u A}}{\a_s^2m\dr{pl}}\)^n
\(\frac n{n+2}\)^{n/2} \(\frac{m_{\u A}}{\a_s}\)^2\cdot \uP\cdot
\(\frac{|\uP|}{\t R(\t\G)}\)^{n/2}.
\eal
\eq79
$$

Using our simplified model for $\RG$, $\uP$ and~$A$ one gets
$$
\al
m_1^2&=\(\frac{m_{\u A}}{\a_s}\)^2 \(\frac{m_{\u A}}{\a_s^2m\dr{pl}}\)^{14}
\cdot 2^{-36}\cr
&\times\frac{\(7g(\z,\mu)+\sqrt{49g^2(\z,\mu)+384h(\z,\mu)}\)^6(\mu^2+4)^8}
{\(\ln\(|\z|+\sqrt{\z^2+1}\)+2\z^2+1\)^6\(2\mu^3+7\mu^2+5\mu+20\)^7}\cr
&\times\(7g(\z,\mu)\sqrt{49g^2(\z,\mu)+384h(\z,\mu)}-98g^2(\z,\mu)
-352h(\z,\mu)\).
\eal
\eq80
$$
Taking $m_{\u A}\simeq m\dr{EW}=80\,$GeV (an electro-weak energy scale) and
$m\dr{pl}\simeq 2.4\cdot 10^{18}\,$GeV, $\a_s^2=\a\dr{em}=\frac1{137}$, one
gets 
$$
m_1\simeq 3.34\cdot 10^{-105}m\dr{EW}G(\z,\mu) \eq81
$$
where
$$
\al
G(\z,\mu)&=\frac{\(7g(\z,\mu)+\sqrt{49g^2(\z,\mu)+384h(\z,\mu)}\)^3(\mu^2+4)^4}
{\(\ln\(|\z|+\sqrt{\z^2+1}\)+2\z^2+1\)^3\(2\mu^3+7\mu^2+5\mu+20\)^{7/2}}\cr
&\times\(7g(\z,\mu)\sqrt{49g^2(\z,\mu)+384h(\z,\mu)}-98g^2(\z,\mu)
-352h(\z,\mu)\)^{1/2}.
\eal
\eq82
$$
In the second de Sitter phase one gets
$$
\al
m_0^2&=\frac14 \(\frac74\)^8 \(\frac{m_{\u A}}{\a_s^2 m\dr{pl}}\)^{14}
\(\frac{m^2_{\u A}}{\a_s^2}\)
\(\frac{\mu^2+4}{2\mu^3+7\mu^2+5\mu+20}\)^7\cr
&\times\(\frac{g(\z,\mu)}{\ln\(|\z|+\sqrt{\z^2+1}\)+2\z^2+1}\)^8. 
\eal \eq83
$$

Taking as before for $m_{\u A}$ an electro-weak energy scale and for $\a_s^2=
\frac1{137}$, one gets
$$
m_0\cong 10^{-102}m\dr{EW}F(\z,\mu) \eq84
$$
where
$$
F(\z,\mu)=\frac{(\mu^2+4)^{7/2}(g(\z,\mu))^4}
{\(\wiel\)^{7/2} \(\lon\)^4}\,. \eq85
$$
According to modern ideas a mass of scalar-\q\ particle should be smaller
than $10^{-33}\,$eV. For the \co\ \ct\ from the second de Sitter phase is the
same for our contemporary epoch, $m_0$~is also a mass of a scalar \q\
particle for our epoch.

It is interesting to ask what is a number of scalar-\q\ particle per unit
volume during both de Sitter phases of the evolution of the Universe (if a
\q\ has been deposed as particles which is not so obvious). (It can be
deposed in a form of a classical scalar field.)
One gets
$$
\al
n_1&=\frac{\rho^1_Q}{m_1}=\frac{6H_0^2}{m_1}\cr
&=\(\frac{m_{\u A}}{\a_s}\)\(\frac{m_{\u A}}{\a^s m\dr{pl}}\)^{n/2}
(2(n+2))^{-(n+4)/4}\cdot 2^5\cr
&\times \frac
{n^2\uP^2-\uP\sqrt{n^2\uP^2+4(n^2-4)A\RG}+4(n+2)A\RG}
{\(2n^2\uP^2+4(n+2)(n-3)A\RG
-n\uP\sqrt{n^2\uP^2+4(n^2-4)A\RG)}\)^{1/2}}\cr
&\times\(-n\uP + \sqrt{n^2\uP^2+4(n^2-4)A\RG}\)^{(n-2)/4}.
\eal \eq86
$$
In the second de Sitter phase one gets
$$
\al
n_0&=\frac{\rho_Q}{m_0}=\frac{6H_1^2}{m_0}\cr
&=\(\frac{m_{\u A}}{m\dr{pl}}\)^{n/2}\(\frac{m_{\u A}}{\a_s^n}\)
\(\frac n{n+2}\)^{(n+4)/4}|\uP|^{1/2}\(\frac{|\uP|}{\RG}\)^{n/4}.
\eal
\eq87
$$
Using our simplified model with $n=14$ one gets from (86)
$$
\al
n_1&
=2^{-19}\(\frac{m_{\u A}}{\a_sm\dr{pl}}\)^7\(\frac{m_{\u A}}{\a^s }\)
\frac{\(7g(\z,\mu)+\sqrt{49g^2(\z,\mu)+384h(\z,\mu)}\)^3}
{\(\wiel\)^{7/2}}\cr
&\times \frac{(\mu^2+4)^3\(98g^2(\z,\mu)+(\mu^2+4)
\sqrt{49g^2(\z,\mu)+384h(\z,\mu)}+32H(\z,\mu)\)}
{\(7g(\z,\mu)\sqrt{49g^2(\z,\mu)+384h(\z,\mu)}-98g^2(\z,\mu)-352h(\z,\mu)\)
^{1/2}}\cr
&\times\(\lon\)^{-1}
\eal
\eq88
$$
and from the formula (87)
$$
\al
n_0&=\(\frac{m_{\u A}}{m\dr{pl}}\)^7\(\frac{m_{\u A}}{\a_s^{14}}\)
\(\frac {7^{9/2}}{2^{17}}\)\cr
&\times\frac{|f(\z)|^{1/2}}{\(\lon\)^{15/2}}
\cdot \(\frac{|f(\z)|(\mu^2+4)^2}{\wiel}\)^7.
\eal
\eq89
$$

For the \co\ \ct\ from the second de Sitter phase is the same for our
contemporary epoch, $n_0$~is also a number of scalar \q\ particles for our
epoch. Taking as usual $m_{\u A}=m\dr{EW}$ and $\a_s^2=\frac1{137}$ one gets
$$
\ealn{
n_1&\simeq 9\cdot 10^{-106}m\dr{EW}\cdot K(\z,\mu) &(90)\cr
n_0&\simeq 6.2 \cdot 10^{-96}m\dr{EW}\cdot L(\z,\mu), &(91)}
$$
where
$$
\ealn{
K(\z,\mu)&=\frac{\(7g(\z,\mu)+\sqrt{49g^2(\z,\mu)+384h(\z,\mu)}\)^3(\mu^2+4)^3}
{\(\wiel\)^{7/2}}\cr
&\times \frac{\(98g^2(\z,\mu)+(\mu^2+4)
\sqrt{49g^2(\z,\mu)+384h(\z,\mu)}+32H(\z,\mu)\)}
{\(7g(\z,\mu)\sqrt{49g^2(\z,\mu)+384h(\z,\mu)}-98g^2(\z,\mu)-352h(\z,\mu)\)
^{1/2}},\qquad &(92)\cr
L(\z,\mu)&=\frac{|f(\z)|^{1/2}}{\(\lon\)^{15/2}}
\cdot \(\frac{|f(\z)|(\mu^2+4)^2}{\wiel}\)^7,&(93)
}
$$
$f(\z)$, $g(\z,\mu)$, $h(\z,\mu)$, $H(\z,\mu)$ are given by the formulae
(40--42). In this way the masses of scalar particles and their numbers per unit
volume in both de Sitter phases and for our contemporary epoch depend on
geometric parameters in our theory.

We can connect $\o \mu$ and $N$ from the formula (90) of Ref.~[3]:
$$
\mu=\frac{5\cdot1.96}{N}\,. \eq94
$$
If $N\simeq60$, one gets
$$
\o\mu=0.16 \eq95
$$
and
$$
5{\o\mu}^2=0.13. \eq96
$$
Using Eq.\ (78) of Ref.~[3] one finds:
$$
n_s(K)=1\pm0.13\,\D\ln K. \eq97
$$

We can also find $\o\mu$ using the formulae (10) and (10a). One gets
$$
\o\mu=\(\frac{m\dr{pl}}{m_{\u A}}\)^2 \a_s^4 F(\b,\g,\a_1)=
\(\frac{m\dr{pl}}{m_{\u A}}\)^4 F(\RG,\uP,A). \eq98
$$
Using Eq.\ (78) (Ref.~[3]) one also gets
$$
n_s(K)=1\pm5\(\frac{m\dr{pl}}{m_{\u A}}\)^8 F^2(\RG,\uP,A)\, \D\ln K. \eq99
$$
Using our simplified model on gets
$$
\al
\o\mu&=81.67\(\frac{m\dr{pl}}{m_{\u A}}\)^4
\frac{7g(\z,\mu)+\sqrt{49g^2(\z,\mu)+384h(\z,\mu)}}
{(\mu^2+4)^2(1+\z^2)}\\
&\times
\frac{|\z|\((5\z^2+4)\ln 2-2\mu^2(1+\z^2)\)}
{\(98g^2(\z,\mu)+(\mu^2+4)\sqrt{49g^2(\z,\mu)+384h(\z,\mu)}+32H(\z,\mu)\)}
\eal
\eq100
$$
and respectively
$$
\al
n_s(K)&=1\pm408.35\,
\frac{\(7g(\z,\mu)+\sqrt{49g^2(\z,\mu)+384h(\z,\mu)}\)^2}
{(\mu^2+4)^4(1+\z^2)^2}\\
&\times
\frac{|\z|^2\((5\z^2+4)\ln 2-2\mu^2(1+\z^2)\)^2\,\D\ln K}
{\(98g^2(\z,\mu)+(\mu^2+4)\sqrt{49g^2(\z,\mu)+384h(\z,\mu)}+32H(\z,\mu)\)^2}\,.
\eal
\eq101
$$

It is easy to see that we can get $n_s(K)$ closed to~1 (a flat power spectral
function) using parameters $\mu$ and~$\z$ in such a way that
$$
(5\z^2+4)\ln2 - 2\mu^2(1+\z^2)
$$
is closed to zero. One can express $\cos\f$ (see Eq.~(5) of Ref.~[3]) in
terms of parameters of the theory and get
$$
\al
\cos\f&=-\(\frac{m_{\u A}}{\a_s^2 m\dr{pl}}\)^{n/2}
\frac{P_1^{n/4}}
{3\cdot2^{(n+8)/4}(n+2)^{n/4}\(\RG\)^{n/4}}\\
&\times\frac{40A\(\dfrac{m_{\u A}}{\a_s^2 m\dr{pl}}\)^n m_{\u A}^2
P_1^{n/2}
+9B\a_s^4 2^{n/2}(n+2)^{n/2}\(\RG\)^{n/2}}
{5A\(\dfrac{m_{\u A}}{\a_s^2 m\dr{pl}}\)^n m_{\u A}^2
P_1^{n/2}
-3\a_s^4 B 2^{n/2}(n+2)^{n/2}\(\RG\)^{n/2}}\,,
\eal
\eq102
$$
where
$$
P_1=-n\uP + \sqrt{n^2\uP^2+4(n^2-4)A\RG}.
$$
For the \ct\ $a$ (see Eq.~(6) of Ref.~[3]) one gets
$$
a=\a_s^2\(\frac{\sqrt B}{m_{\u A}}\)\(\frac{\a_s^2 m\dr{pl}}{m_{\u A}}\)^{n/2}
\frac{2^{n/4}(n+2)^{n/4}\(\RG\)^{n/4}}
{\sqrt A\(-n\uP + \sqrt{n^2\uP^2+4(n^2-4)A\RG}\)^{n/4}}.
\eq103
$$

\def\ii#1 {\item{[#1]}}
\section{References}
\setbox0=\hbox{[11]\enspace}
\parindent\wd0

\ii1 {Kalinowski} M. W., 
{\it Dynamics of Higgs' field and a quintessence
in the nonsymmetric Kaluza--Klein (Jordan--Thiry) theory}, 
arXiv: hep-th/0306241v1. 

\ii2 {Kalinowski} M. W., 
{\it Nonsymmetric Fields Theory and its Applications\/},
World Scientific, Singapore, New Jersey, London, Hong Kong 1990.

\item{} {Kalinowski} M. W., 
{\it Nonsymmetric Kaluza--Klein (Jordan--Thiry) Theory in a general
nonabelian case}\/,
Int. Journal of Theor. Phys. {\bf30}, p.~281 (1991).

\item{} {Kalinowski} M. W., 
{\it Nonsymmetric Kaluza--Klein (Jordan--Thiry) Theory in the electromagnetic
case}\/,
Int. Journal of Theor. Phys. {\bf31}. p.~611 (1992).

\item{} {Kalinowski} M. W., 
{\it Can we get confinement from extra dimensions}\/,
in: Physics of Elementary Interactions (ed. Z.~Ajduk, S.~Pokorski,
A.~K.~Wr\'oblewski), World Scientific, Singapore, New
Jersey, London, Hong Kong 1991.

\item{} {Kalinowski} M. W., 
{\it Scalar fields in the nonsymmetric Kaluza--Klein (Jordan--Thiry)
theory}, arXiv: hep-th/0307242.

\ii3 {Kalinowski} M. W., 
{\it On some \co\ consequences of the Nonsymmetric\break
Kaluza--Klein (Jordan--Thiry) Theory\/}, arXiv: hep-th/0307289.

\ii4 {Manton} N. S.,
{\it A new six-dimensional approach to the Weinberg-Salam model\/},
Nucl. Phys. {\bf B158}, p.~141 (1979).

\end